\documentclass[journal]{IEEEtran}
\ifCLASSINFOpdf
\else
   \usepackage{graphicx}
   \graphicspath{{../eps/}}
   \DeclareGraphicsExtensions{.eps}
\fi

\usepackage{amsmath}
\usepackage{amsfonts,amssymb}
\usepackage{amssymb}
\usepackage{wrapfig}
\usepackage{psfrag}
\usepackage{epstopdf}
\usepackage{cite}
\usepackage{graphicx}
\usepackage{subfigure}
\usepackage{threeparttable}
\usepackage{cases}
\usepackage{subeqnarray}
\usepackage{color}
\usepackage{underscore}
\usepackage{verbatim}
\usepackage{bm}
\usepackage{stfloats}
\usepackage{xpatch}

\newtheorem{algorithm}{Algorithm}

\usepackage{algorithm}
\usepackage{algorithmic}

\makeatletter

\begin{document}
\title{Group Movable Antenna With Flexible Sparsity: Joint Array Position and Sparsity Optimization}
%
%
%
\author{
        Haiquan~Lu,
        Yong~Zeng,~\IEEEmembership{Senior Member,~IEEE,}
        Shi~Jin,~\IEEEmembership{Fellow,~IEEE,}
        and
        Rui~Zhang,~\IEEEmembership{Fellow,~IEEE}
\thanks{This work was supported in part by the Natural Science Foundation for Distinguished Young Scholars of Jiangsu Province under Grant BK20240070, and in part by the National Natural Science Foundation of China under Grant 62071114. (\emph{Corresponding author: Yong Zeng.}) }
\thanks{Haiquan Lu, Yong Zeng, and Shi Jin are with the National Mobile Communications Research Laboratory and Frontiers Science Center for Mobile Information Communication and Security, Southeast University, Nanjing 210096, China. Haiquan Lu and Yong Zeng are also with the Purple Mountain Laboratories, Nanjing 211111, China (e-mail: \{haiquanlu, yong_zeng, jinshi\}@seu.edu.cn). }
\thanks{Rui Zhang is with School of Science and Engineering, Shenzhen Research Institute of Big Data, The Chinese University of Hong Kong, Shenzhen, Guangdong 518172, China (e-mail: rzhang@cuhk.edu.cn). He is also with the Department of Electrical and Computer Engineering, National University of Singapore, Singapore 117583 (e-mail: elezhang@nus.edu.sg).}
}

\maketitle

\begin{abstract}
 Movable antenna (MA) is a promising technology to exploit the spatial variation of wireless channel for performance enhancement, by dynamically varying the antenna position within a certain region. However, for multi-antenna communication systems, moving each antenna independently not only requires prohibitive complexity to find the optimal antenna positions, but also incurs sophisticated movement control in practice. To address this issue, this letter proposes a new MA architecture termed \emph{group MA} (GMA), enabling the group movement of all elements collectively in a continuous manner, and simultaneously achieving flexible array architecture by antenna selection (AS). In this letter, we focus on the uniform sparse array based GMA, where equally spaced antenna elements are selected to achieve desired array sparsity. The array position and sparsity level are jointly optimized to maximize the sum rate of the multi-user communication system. Numerical results verify the necessity to optimize the position and sparsity of GMA, and considerable performance gain is achieved as compared to the conventional fixed-position antenna (FPA).
\end{abstract}
%

\IEEEpeerreviewmaketitle
\vspace{-0.6cm}
\section{Introduction}
 The recent evolution of wireless communication networks is accompanied by the advancement of multi-antenna technology, e.g., from multiple-input multiple-output (MIMO) in the fourth-generation (4G) to massive MIMO in the fifth-generation (5G) networks. Looking forward towards the future sixth-generation (6G), massive MIMO is expected to evolve towards extremely large-scale MIMO (XL-MIMO) \cite{lu2024tutorial}. Despite achieving higher spatial resolution and spectral efficiency, conventional multi-antenna technologies mainly utilize the fixed-position antenna (FPA) architecture with adjacent array elements typically separated by half-wavelength, termed as compact array. In this case, the ever-increasing antenna size is faced with practical issues of high hardware cost and energy expenditure. To address these issues, one effective approach is antenna selection (AS), where only a subset of array elements with favourable channel conditions are activated to harness the diversity/multiplexing gain \cite{sanayei2004antenna}. Another promising line of research is to design flexible array architectures by pre-configuring the antenna spacing, e.g., modular and sparse array \cite{li2024sparse,wang2023can}, which is able to achieve larger array aperture than the compact array with the same number of array elements. Thus, a narrower beamwidth of the main lobe is enabled for better interference mitigation, rendering it quite appealing for hot-spot areas with densely located users \cite{wang2023can}. Besides, sparse array has been designed for exploiting enhanced spatial resolution and degree-of-freedom (DoF) in radar sensing systems, thanks to its enlarged virtual aperture \cite{pal2010nested}, thus increasing the number of targets that can be distinguished simultaneously.

 Recently, movable antennas (MAs) \cite{zhu2024movable,ma2024mimo,yang2024flexible,mei2024movable} or fluid antenna systems (FASs) \cite{wong2021fluid,wang2024fluid,ghadi2024cache} have been proposed as a new approach to fully exploit the channel spatial variations. Specifically, driven by step motors or servos \cite{basbug2017design}, each MA is connected to the radio frequency (RF) chain through a flexible cable, and is thus endowed with the capability of free movement \cite{zhu2024movable,zhang2024movable}. Compared to contemporary massive MIMO systems with FPAs, MAs can flexibly adjust each antenna position in a continuous manner within certain region \cite{zeng2024fixed}, so as to achieve capacity enhancement \cite{zhu2024movable,ma2024mimo}, interference mitigation \cite{zhu2023movable}, secure transmission \cite{hu2024secure}, integrated sensing and communication (ISAC) enhancement \cite{lyu2024flexible,zeng2024fixed,wang2024fluid}, and quality of service (QoS) improvement \cite{ghadi2024cache}, etc. Besides, the prototype and experimental results of MA were reported in  \cite{dong2024movable}. However, for multi-antenna systems, the independent movement of each antenna not only results in prohibitive complexity to find the optimal antenna positions, but also incurs sophisticated movement control, which hinders its practical deployment.

 To tackle the above issues, in this letter, we propose a new MA architecture termed \emph{group MA} (GMA). Instead of moving each antenna element independently, the entire antenna array is moved collectively in a continuous manner. Moreover, a flexible array architecture, such as compact/modular/sparse array, can be achieved by activating/deactivating the corresponding array elements by AS. Such a GMA architecture not only achieves the continuous movement in the spatial domain, but also significantly reduces the implementation complexity, without having to adjust each antenna position individually, thus providing a feasible solution to the practical deployment of MAs. The main contributions of this work are summarized as follows. First, we propose a new GMA architecture for ease of the practical deployment. Second, by focusing on the uniform sparse array based GMA, where the antenna spacing is determined by \emph{sparsity} that is typically larger than half signal wavelength, the position and sparsity level of the GMA are jointly optimized to maximize the sum rate of users. An alternating optimization algorithm is proposed to tackle the non-convex optimization problem. Numerical results demonstrate the considerable performance gain of the GMA, and the necessity of joint position and sparsity optimization for GMA.

\vspace{-0.6cm}
\section{System Model and Problem Formulation}\label{sectionSystemModel}

 \begin{figure}[!t]
 \setlength{\abovecaptionskip}{-0.2cm}
 \setlength{\belowcaptionskip}{-0.1cm}
 \centering
 \centerline{\includegraphics[width=2.5in,height=2.0in]{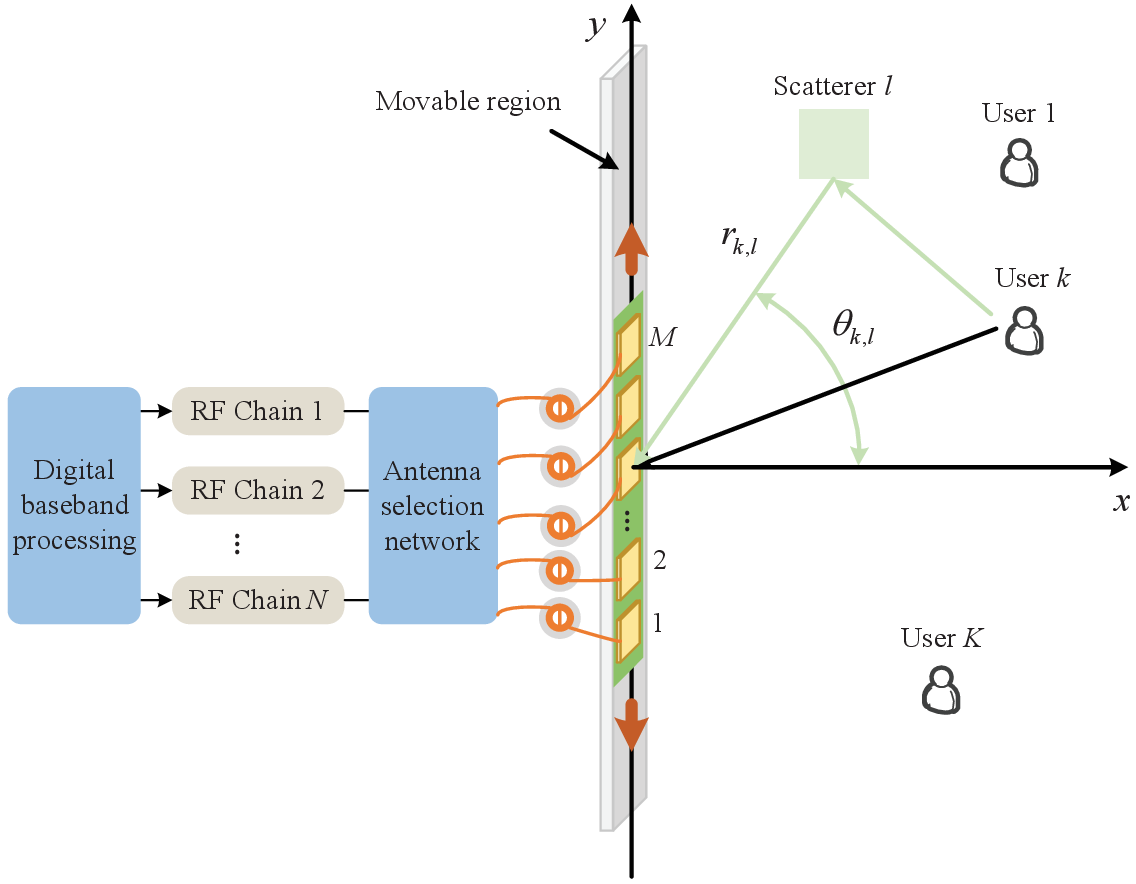}}
 \caption{A GMA-enabled multi-user communication system with flexible array sparsity.}
 \label{fig:systemModel}
 \vspace{-0.7cm}
 \end{figure}
 As shown in Fig.~\ref{fig:systemModel}, we consider a GMA-enabled multi-user communication system, where $K$ single FPA users are served by a GMA.{\footnote[1]{ Due to the practical considerations for space and power supply, the MA is only considered at the base station (BS). The extension of the MA to users is worthy of investigation in future work.}} The GMA consists of a physical array with $M$ elements that are separated by $d=\lambda/2$, with $\lambda$ being the signal wavelength, and the number of RF chains is $N < M$. Besides, by activating/deactivating proper elements of the array, a uniform sparse array of $N$ elements is formed, where adjacent elements are separated by $\eta d$, with $\eta \ge 1$ denoting the sparsity level \cite{wang2023can}. The dimension of the physical array is $D = \left( {M- 1} \right) d$, and that of the resulting uniform sparse array after AS is ${D_s} = \left( {N - 1} \right)\eta d$. Without loss of generality, let the bottom element be the reference point of the uniform sparse array. Under the two-dimensional (2D) Cartesian coordinate system, the coordinate of the reference point is denoted as ${\bf{w}} = {\left[ {0,y} \right]^T} \in {\cal C}$, where $\cal C$ denotes the freely movable region. Thus, the position of array element $n$ of the uniform sparse array is given by ${{\bf{w}}_n} = {\bf{w}} + {\left[ {0,\left( {n - 1} \right)\eta d} \right]^T}$. Let $L_k$ denote the number of scatterers experienced by user $k$, and ${{\bf{q}}_{k,l}} = {\left[ {{r_{k,l}}\cos {\theta _{k,l}},{r_{k,l}}\sin {\theta _{k,l}}} \right]^T}$ denote the position of scatterer $l$ for user $k$, with ${{r_{k,l}}}$ being the distance between the origin ${\bf o}$ and scatterer $l$ of user $k$, and $\theta_{k,l}$ being the angle of arrival (AoA). Note that for the line-of-sight (LoS) channel, the scatterer can be treated as the user itself.

 For any given AoA $\theta$, the normalized wave propagation direction vector is ${\bm{\kappa}} = {\left[ {\cos \theta ,\sin \theta } \right]^T}$, and the distance difference of the signal propagation between the position $\bf w$ and the origin is $\rho \left( {y} \right) = {{\bm{\kappa}}^T}\left( {{\bf{w}} - {\bf{o}}} \right) = y\sin \theta $. Since the GMA moves along the $y$-axis, the notation $y$ is used for denoting the reference position of the GMA in the following. Thus, the receive array response vector is a function of the position $y$, sparsity level $\eta$, and the AoA $\theta$, given by
 \begin{equation}\label{arrayResponseVector}
 {\bf{a}}\left( {{y},\eta ;\theta } \right) = {e^{{\rm j}\frac{{2\pi }}{\lambda }\rho \left( {y} \right)}}{\bf{\bar a}}\left( {\eta ;\theta } \right),
 \end{equation}
 where ${\rm j}$ denotes the imaginary unit of complex numbers, with ${{\rm j}^2} =  - 1$, and ${\bf{\bar a}}\left({\eta ;\theta } \right) \in {\mathbb C}^{N \times 1}$ represents the array response vector of the resulting uniform sparse array, given by \cite{wang2023can}
 \begin{equation}
 {\bf{\bar a}}\left( {\eta ;\theta } \right) = {\left[ {1, \cdots ,{e^{{\rm j}2\pi \left( {n - 1} \right)\eta \bar d\sin \theta }}, \cdots ,{e^{{\rm j}2\pi \left( {N - 1} \right)\eta \bar d\sin \theta }}} \right]^T},
 \end{equation}
 where ${\bar d} \triangleq d/\lambda$.

 We consider the quasi-static slow fading channels, and the channel from user $k$ to the GMA is
 \begin{equation}
 \begin{aligned}
 &{{\bf{h}}_k}\left( {{y},\eta } \right) = \sum\limits_{l = 1}^{{L_k}} {{\alpha _{k,l}}{\bf{a}}\left( {{y},\eta ;{\theta _{k,l}}} \right)}  = \sum\limits_{l = 1}^{{L_k}} {{\alpha _{k,l}}{e^{{\rm j}\frac{{2\pi }}{\lambda }y\sin {\theta _{k,l}}}} \times } \\
 &\ \ \ {\left[ {1, \cdots ,{e^{{\rm j}2\pi \left( {n - 1} \right)\eta \bar d\sin {\theta _{k,l}}}}, \cdots ,{e^{{\rm j}2\pi \left( {N - 1} \right)\eta \bar d\sin {\theta _{k,l}}}}} \right]^T},
 \end{aligned}
 \end{equation}
 where ${\alpha _{k,l}}$ denotes the complex-valued path gain of path $l$ for user $k$. To reveal the fundamental performance limit, perfect channel state information (CSI) is assumed at the GMA.

 To detect the symbol for user $k$, the receive beamforming/combining vector ${{\bf v}_k} \in {\mathbb C}^{N \times 1}$ is applied, where $\left\| {{{\bf{v}}_k}} \right\| = 1$. The resulting signal at the GMA is
 \begin{equation}\label{resultingSignal}
 \begin{aligned}
 z\left( {{y},\eta } \right) =  \ & {\bf{v}}_k^H{{\bf{h}}_k}\left( {{y},\eta } \right)\sqrt {{P_k}} {s_k} + \\
 &{\bf{v}}_k^H\sum\limits_{i = 1,i \ne k}^K {{{\bf{h}}_i}\left( {{y},\eta } \right)\sqrt {{P_i}} {s_i}}  + {\bf{v}}_k^H{\bf{n}},
 \end{aligned}
 \end{equation}
 where $s_i$ and $P_i$ represent the information-bearing symbol and transmit power of user $i$, respectively, $1 \le i\le K$, ${\bf{n}} \sim {\cal CN}\left( {{\bf{0}},{\sigma ^2}{{\bf{I}}_N}} \right)$ represents the complex-valued additive white Gaussian noise (AWGN) with zero mean and average power $\sigma ^2$. The resulting signal-to-interference-plus-noise ratio (SINR) for user $k$ is
 \begin{equation}\label{SINRuserk}
 \begin{aligned}
 {\gamma _k}\left( {{y},\eta } \right) &= \frac{{{{\bar P}_k}{{\left| {{\bf{v}}_k^H{{\bf{h}}_k}\left( {{y},\eta } \right)} \right|}^2}}}{{\sum\limits_{i = 1,i \ne k}^K {{{\bar P}_i}{{\left| {{\bf{v}}_k^H{{\bf{h}}_i}\left( {{y},\eta } \right)} \right|}^2}}  + 1}}\\
 & = {\bar P_k}\frac{{{\bf{v}}_k^H{{\bf{h}}_k}\left( {{y},\eta } \right){\bf{h}}_k^H\left( {{y},\eta } \right){{\bf{v}}_k}}}{{{\bf{v}}_k^H{{\bf{C}}_k}\left( {{y},\eta } \right){{\bf{v}}_k}}},
 \end{aligned}
 \end{equation}
 where ${{\bar P}_i} \triangleq {P_i}/\sigma^2$, and ${{\bf{C}}_k}\left( {{y},\eta } \right) \triangleq {\bf{I}} + \sum\nolimits_{i = 1,i \ne k}^K {{{\bar P}_i}{{\bf{h}}_i}\left( {{y},\eta } \right){\bf{h}}_i^H\left( {{y},\eta } \right)} $ represents the interference-plus-noise covariance matrix of user $k$.

 A closer look at \eqref{SINRuserk} reveals that ${\gamma _k}\left( {{y},\eta } \right)$ is the generalized Rayleigh quotient of ${\bf v}_k$, and thus, the optimal solution for ${\bf v}_k$ is given by ${\bf{v}}_k^ \star  = {\bf{C}}_k^{ - 1}\left( {y,\eta } \right){{\bf{h}}_k}\left( {y,\eta } \right)/$ $\left\| {{\bf{C}}_k^{ - 1}\left( {y,\eta } \right){{\bf{h}}_k}\left( {y,\eta } \right)} \right\|$, yielding the resulting SINR given by
 \begin{equation}\label{reducedSINRuserk}
 {\gamma _k}\left( {{y},\eta } \right) = {{\bar P}_k}{\bf{h}}_k^H\left( {{y},\eta } \right){\bf{C}}_k^{ - 1}\left( {{y},\eta } \right){{\bf{h}}_k}\left( {{y},\eta } \right).
 \end{equation}
 The achievable sum rate of all users is thus given by
 \begin{equation}\label{sumRate}
 R\left( {{y},\eta } \right) = \sum\limits_{k = 1}^K {{{\log }_2}\left( {1 + {\gamma _k}\left( {{y},\eta } \right)} \right)}.
 \end{equation}

 Our objective is to maximize the achievable rate by jointly optimizing the position $y$ and sparsity level $\eta$ of the GMA. The optimization problem can be formulated as
 \begin{equation}
 \begin{aligned}
  \left( {\rm{P1}} \right)\ \mathop {\max }\limits_{{y},\eta }&\ \ R\left( {{y},\eta } \right)\\
 {\rm{s.t.}}&\ \ \eta  \in \left\{ {1, \cdots ,{\eta _{\max }}} \right\},\\
 &\ \ {y_{\min }} \le y \le {y_{\max }},
 \end{aligned}
 \end{equation}
 where ${\eta _{\max }} \triangleq \left\lfloor {(M - 1)/(N - 1)} \right\rfloor $ denotes the maximum sparsity level, which stems from the fact that the array dimension of the resulting uniform sparse array with AS cannot exceed that of the original physical array, and $y_{\min}$ and $y_{\max}$ represent the minimum and maximum movable positions along the $y$-axis, respectively.

\section{Proposed Solution}\label{section}
 In this section, we propose an alternating optimization algorithm to solve problem (P1). To gain useful insights, we first consider the special case of the single-user communication.

 \subsection{Single-User Communication}
 For the single-user communication, the resulting signal in \eqref{resultingSignal} reduces to
 \begin{equation}\label{resultingSignalSingleuser}
 z\left( {{y},\eta } \right) = {{\bf{v}}^H}{\bf{h}}\left( {{y},\eta } \right)\sqrt P s + {{\bf{v}}^H}{\bf{n}},
 \end{equation}
 where the user index is omitted for brevity. It is known that the maximal-ratio combining (MRC) beamforming is optimal, i.e., ${\bf{v}}^{\star} = {\bf{h}}\left( {{y},\eta } \right)/\left\| {{\bf{h}}\left( {{y},\eta } \right)} \right\|$, and the resulting signal-to-noise ratio (SNR) is
 \begin{equation}\label{resultingSNR}
 \begin{aligned}
 \gamma \left( {{y},\eta } \right) &= \bar P{\left\| {{\bf{h}}\left( {{y},\eta } \right)} \right\|^2} = \bar P{\left\| {\sum\limits_{l = 1}^L {{\alpha _l}{\bf{a}}\left( {{y},\eta ;{\theta _l}} \right)} } \right\|^2}\\
 &= \bar P {\left\| {{\bf{A}}\left( \eta  \right){\bf{f}}\left( {y} \right)} \right\|^2},
 \end{aligned}
 \end{equation}
 where ${\bf{A}}\left( \eta  \right) \in {{\mathbb C}^{N \times L}} \triangleq \left[ {{\alpha _1}{\bf{\bar a}}\left( {\eta ;{\theta _1}} \right), \cdots ,{\alpha _L}{\bf{\bar a}}\left( {\eta ;{\theta _L}} \right)} \right]$ and ${\bf{f}}\left( {y} \right) \in {{\mathbb C}^{L \times 1}} = {[ {{e^{{\rm j}\frac{{2\pi }}{\lambda }y\sin {\theta _1}}}, \cdots ,{e^{{\rm j}\frac{{2\pi }}{\lambda }y\sin {\theta _L}}}} ]^T}$.

 The SNR maximization problem can be equivalently formulated as
 \begin{equation}\label{SNRMaximizationProblem}
 \begin{aligned}
 \mathop {\max }\limits_{{y},\eta }&\ \ {\left\| {{\bf{A}}\left( \eta  \right){\bf{f}}\left( {y} \right)} \right\|^2}\\
 {\rm{s.t.}}&\ \ \eta  \in \left\{ {1, \cdots ,{\eta _{\max }}} \right\},\\
 &\ \ {y_{\min }} \le y \le {y_{\max }}.
 \end{aligned}
 \end{equation}

 Note that problem \eqref{SNRMaximizationProblem} is challenging to be directly solved since the problem is non-convex, and the optimization variables $y$ and $\eta$ are coupled with each other in the objective function. In the following, we propose an alternating optimization algorithm to solve \eqref{SNRMaximizationProblem}.

 \subsubsection{Optimization of $y$ with Given $\eta$}
 For any given array sparsity level $\eta$, the subproblem of optimizing the position $y$ to maximize the SNR is written as
 \begin{equation}\label{equivalentSNRMaximizationProblem}
 \begin{aligned}
 \mathop {\max }\limits_{y }&\ \ {\left\| {{\bf{A}}\left( \eta  \right){\bf{f}}\left( y \right)} \right\|^2}\\
 {\rm{s.t.}}&\ \ {y_{\min }} \le y \le {y_{\max }}.
 \end{aligned}
 \end{equation}
 To tackle the non-convex problem, the successive convex approximation (SCA) is applied, which is an iterative optimization technique that updates the optimization variable over each iteration \cite{zeng2019accessing}.

 First, a closer look at the objective function reveals that ${\left\| {{\bf{A}}\left( \eta  \right){\bf{f}}\left( y \right)} \right\|^2}$ is convex with respect to ${{\bf{f}}\left( y \right)}$, whose first-order Taylor approximation is a global under-estimator \cite{zeng2019accessing}, given by
 \begin{equation}
 \begin{aligned}
 &{\left\| {{\bf{A}}\left( \eta  \right){\bf{f}}\left( y \right)} \right\|^2} \ge {\left\| {{\bf{A}}\left( \eta  \right){\bf{f}}\left( {{y^{\left( j \right)}}} \right)} \right\|^2} +  \\
 &\ \ \ 2{\rm{Re}}\left\{ {{{\bf{f}}^H}\left( {{y^{\left( j \right)}}} \right){{\bf{A}}^H}\left( \eta  \right){\bf{A}}\left( \eta  \right)\left( {{\bf{f}}\left( y \right) - {\bf{f}}\left( {{y^{\left( j \right)}}} \right)} \right)} \right\}\\
 &\ \ \ = 2g\left( y \right) - {\left\| {{\bf{A}}\left( \eta  \right){\bf{f}}\left( {{y^{\left( j \right)}}} \right)} \right\|^2},
 \end{aligned}
 \end{equation}
 where ${y^{\left( j \right)}}$ represents the resulting position in the $j$-th iteration, $g\left( y \right) \triangleq {\mathop{\rm Re}\nolimits} \left\{ {{{\bf{f}}^H}\left( {{y^{\left( j \right)}}} \right){{\bf{A}}^H}\left( \eta  \right){\bf{A}}\left( \eta  \right){\bf{f}}\left( y \right)} \right\}$, and ${\mathop{\rm Re}\nolimits} \left\{ {\cdot} \right\}$ denotes the real part of a complex number. Though $g\left( y \right)$ is linear with respect to ${\bf{f}}\left( y \right)$, it is neither convex nor concave with respect to $y$. To this end, by applying the second-order Taylor approximation, a quadratic surrogate function is constructed to serve as a global lower bound of $g\left( y \right)$. Specifically, let ${\bf{b}} \in {{\mathbb C}^{L \times 1}} \triangleq {{\bf{A}}^H}\left( \eta  \right){\bf{A}}\left( \eta  \right){\bf{f}}\left( {{y^{\left( j \right)}}} \right)$, and its $i$-th entry is denoted as ${b_i} = \left| {{b_i}} \right|{e^{{\rm j}\angle {b_i}}}$, with $\left| {{b_i}} \right|$ and ${\angle {b_i}}$ being the amplitude and phase, respectively, $1 \le i \le L$. Thus, $g\left( y \right)$ can be expressed as
 \begin{equation}
 \begin{aligned}
 g\left( y \right) &= {\mathop{\rm Re}\nolimits} \left\{ {\sum\limits_{i = 1}^L {\left| {{b_i}} \right|{e^{{\rm j}\left( {\frac{{2\pi }}{\lambda }y\sin {\theta _i} - \angle {b_i}} \right)}}} } \right\}\\
 &= \sum\limits_{i = 1}^L {\left| {{b_i}} \right|\cos \left( {\frac{{2\pi }}{\lambda }y\sin {\theta _i} - \angle {b_i}} \right)}.
 \end{aligned}
 \end{equation}

 The first- and second-order derivatives of $g\left( y \right)$ over ${y^{\left( j \right)}}$ are respectively given by
 \begin{equation}\label{firstOrderDerivative}
 g'\left( {{y^{\left( j \right)}}} \right) =  - \frac{{2\pi }}{\lambda }\sum\limits_{i = 1}^L {\left| {{b_i}} \right|\sin {\theta _i}\sin \left( {\frac{{2\pi }}{\lambda }{y^{\left( j \right)}}\sin {\theta _i} - \angle {b_i}} \right)},
 \end{equation}
 \begin{equation}\label{secondOrderDerivative}
 g''\left( {{y^{\left( j \right)}}} \right) =  - \frac{{4{\pi ^2}}}{{{\lambda ^2}}}\sum\limits_{i = 1}^L {\left| {{b_i}} \right|{{\sin }^2}{\theta _i}\cos \left( {\frac{{2\pi }}{\lambda }{y^{\left( j \right)}}\sin {\theta _i} - \angle {b_i}} \right)}.
 \end{equation}
 It then follows that $g''\left( {{y^{\left( j \right)}}} \right) \le \xi$, with $\xi  \triangleq \frac{{4{\pi ^2}}}{{{\lambda ^2}}}\sum\nolimits_{i = 1}^L {\left| {{b_i}} \right|} > 0$. The quadratic surrogate function is then given by \cite{ma2024mimo}
 \begin{equation}
 \begin{aligned}
 g\left( y \right) &\ge g\left( {{y^{\left( j \right)}}} \right) + g'\left( {{y^{\left( j \right)}}} \right)\left( {y - {y^{\left( j \right)}}} \right) - \frac{\xi}{2} {\left( {y - {y^{\left( j \right)}}} \right)^2}\\
 &= \underbrace { - \frac{1}{2}\xi {y^2} + \left( {g'\left( {{y^{\left( j \right)}}} \right) + \xi {y^{\left( j \right)}}} \right)y}_{\bar g\left( y \right)} + \\
 &\ \ \ \ \ g\left( {{y^{\left( j \right)}}} \right) - g'\left( {{y^{\left( j \right)}}} \right){y^{\left( j \right)}} - \frac{1}{2}\xi {\left( {{y^{\left( j \right)}}} \right)^2}.
 \end{aligned}
 \end{equation}

 Thus, for given ${y^{\left( j \right)}}$, problem \eqref{equivalentSNRMaximizationProblem} is lower-bounded by the following problem,
 \begin{equation}\label{lowBoundSNRMaximizationProblem}
 \begin{aligned}
 \mathop {\max }\limits_{y }&\ \ { - \frac{1}{2}\xi {y^2} + \left( {g'\left( {{y^{\left( j \right)}}} \right) + \xi {y^{\left( j \right)}}} \right)y}\\
 {\rm{s.t.}}&\ \ {y_{\min }} \le y \le {y_{\max }}.
 \end{aligned}
 \end{equation}
 It is observed that the objective function is a concave function of $y$, and the optimal solution to \eqref{lowBoundSNRMaximizationProblem} is given by
 \begin{equation}
 {y^{\left( {j + 1} \right)}} = \left\{ \begin{split}
 &{y_{\min}},\ {\rm{if }}\ {y_{\min }}{\rm{ > }}\frac{1}{\xi }g'\left( {{y^{\left( j \right)}}} \right) + {y^{\left( j \right)}},\\
 &\frac{1}{\xi }g'\left( {{y^{\left( j \right)}}} \right) + {y^{\left( j \right)}},\\
 &\ \ \ \ \ \ \ \ {\rm{if }}\ {y_{\min }} \le \frac{1}{\xi }g'\left( {{y^{\left( j \right)}}} \right) + {y^{\left( j \right)}} \le {y_{\max }},\\
 &{{y}_{\max }},\ {\rm{ if }}\ {y_{\max }} < \frac{1}{\xi }g'\left( {{y^{\left( j \right)}}} \right) + {y^{\left( j \right)}}.
 \end{split} \right.
 \end{equation}
 Then, the optimized position for given $\eta$ can be obtained by iteratively solving \eqref{lowBoundSNRMaximizationProblem}. Note that the objective function is non-decreasing over each iteration, and thus the convergence of the iterative algorithm is guaranteed.

 \subsubsection{Optimization of $\eta$ with Given $y$}
 For given position $y$, the subproblem of optimizing the sparsity level $\eta$ to maximize the SNR is written as
 \begin{equation}\label{equivalentSNRMaximizationProblemEta}
 \begin{aligned}
 \mathop {\max }\limits_{\eta }&\ \ {\left\| {{\bf{A}}\left( \eta  \right){\bf{f}}\left( {y} \right)} \right\|^2}\\
 {\rm{s.t.}}&\ \ \eta  \in \left\{ {1, \cdots ,{\eta _{\max }}} \right\}.\\
 \end{aligned}
 \end{equation}
 Since the sparsity level $\eta$ intricately appears in each entry of ${\bf{A}}\left( \eta  \right)$, which is difficult to be directly optimized. Fortunately, $\eta$ only has finite discrete values, and its optimal value can be efficiently obtained via traversing the finite candidate values.

 As a result, problem \eqref{SNRMaximizationProblem} can be solved by iteratively optimizing the position $y$ and sparsity level $\eta$, until the solution  convergence is achieved. The main procedure of this solution is summarized in Algorithm~\ref{alg1}. Regarding the complexity of Algorithm~\ref{alg1}, step 3 has the complexity of  ${\rm{{\cal O}}}({I_1})$, where $I_1$ denotes the total number of iterations required by SCA to converge. The complexity for step 4 is ${\rm{{\cal O}}}({\eta _{\max }})$. Thus, the overall complexity of  Algorithm~\ref{alg1} is ${\rm{{\cal O}}}({I_2}{I_1} + {I_2}{\eta _{\max }})$, where $I_2$ denotes the total number of iterations required for Algorithm~\ref{alg1} to converge.

 \begin{algorithm}[t]
 \caption{Alternating Optimization for Problem \eqref{SNRMaximizationProblem}}
 \label{alg1}
 \begin{algorithmic}[1]
 \STATE Initialize ${\eta ^{\left( 0 \right)}}$, and let $r=0$.
 \STATE \textbf{repeat}
 \STATE For given ${\eta ^{\left( r \right)}}$, solve problem \eqref{lowBoundSNRMaximizationProblem} via SCA technique, and denote the solution as ${y^{\left( {r + 1} \right)}}$.
 \STATE For given ${y^{\left( {r + 1} \right)}}$, obtain the optimal solution to \eqref{equivalentSNRMaximizationProblemEta} via a discrete search, and denote the solution as ${\eta ^{\left( {r + 1} \right)}}$.
 \STATE Update $r = r+1$.
 \STATE \textbf{until} the fractional increase of the objective function is below
 a given threshold $\epsilon >0$.
 \end{algorithmic}
 \end{algorithm}

 \vspace{-0.5cm}
 \subsection{Multi-User Communication}
 In this subsection, we consider the more general multi-user communication, and problem (P1) is solved via the alternating optimization technique. Specifically, for given sparsity level $\eta$, the subproblem of optimizing the position is given by
 \begin{equation}\label{subProblemMultiuserLocation}
 \begin{aligned}
 \mathop {\max }\limits_{y }&\ \ R\left( {y,\eta } \right)\\
 {\rm{s.t.}}&\ \ {y_{\min}} \le y \le {y_{\max}}.\\
 \end{aligned}
 \end{equation}
 Besides, for given position $y$, the subproblem of optimizing the sparsity level is
 \begin{equation}\label{subProblemMultiuserSparsityParameter}
 \begin{aligned}
 \mathop {\max }\limits_{\eta }&\ \ R\left( {y,\eta } \right)\\
 {\rm{s.t.}}&\ \  \eta  \in \left\{ {1, \cdots ,{\eta _{\max }}} \right\}.\\
 \end{aligned}
 \end{equation}

 Since the position and sparsity level optimization need to balance different paths of all users, instead of resorting to the sophisticated optimization methods, the one-dimensional continuous search and discrete search are respectively applied for obtaining the optimal position and sparsity level in \eqref{subProblemMultiuserLocation} and \eqref{subProblemMultiuserSparsityParameter}. Similarly, problem (P1) can be solved by iteratively optimizing $y$ and $\eta$, until the convergence is reached.

 \vspace{-0.3cm}
\section{Numerical Results}\label{sectionNumericalResults}
 In this section, numerical results are provided to validate the performance of the proposed GMA system. We consider the millimeter wave (mmWave) system that operates at the carrier frequency $f = 28$ GHz. The number of users is $K = 5$, which are uniformly distributed in the circular area with the center and radius given by ${{\bf{q}}_c} = {\left[ {100,0} \right]^T}$ m and $r_{\rm radius} = 50$ m. The number of scatterers experienced by each user is $L_k = 5$, which are uniformly distributed in ${r_{k,l}} \in \left[ {0,75} \right]$ m and ${\theta _{k,l}} \in \left[ { - \pi /2,\pi /2} \right]$. The transmit power of each user is $P_k = 10$ dBm, and the noise power spectrum density is $N_0 = -174$ dBm/Hz. The number of RF chains is $N=4$.

 \begin{figure}
 \setlength{\abovecaptionskip}{-0.1cm}
 \setlength{\belowcaptionskip}{-0.1cm}
 \centering
 \subfigure[SNR of the single-user communication]{
 \begin{minipage}[t]{0.5\textwidth}
 \centering
 \centerline{\includegraphics[width=2.5in,height=1.875in]{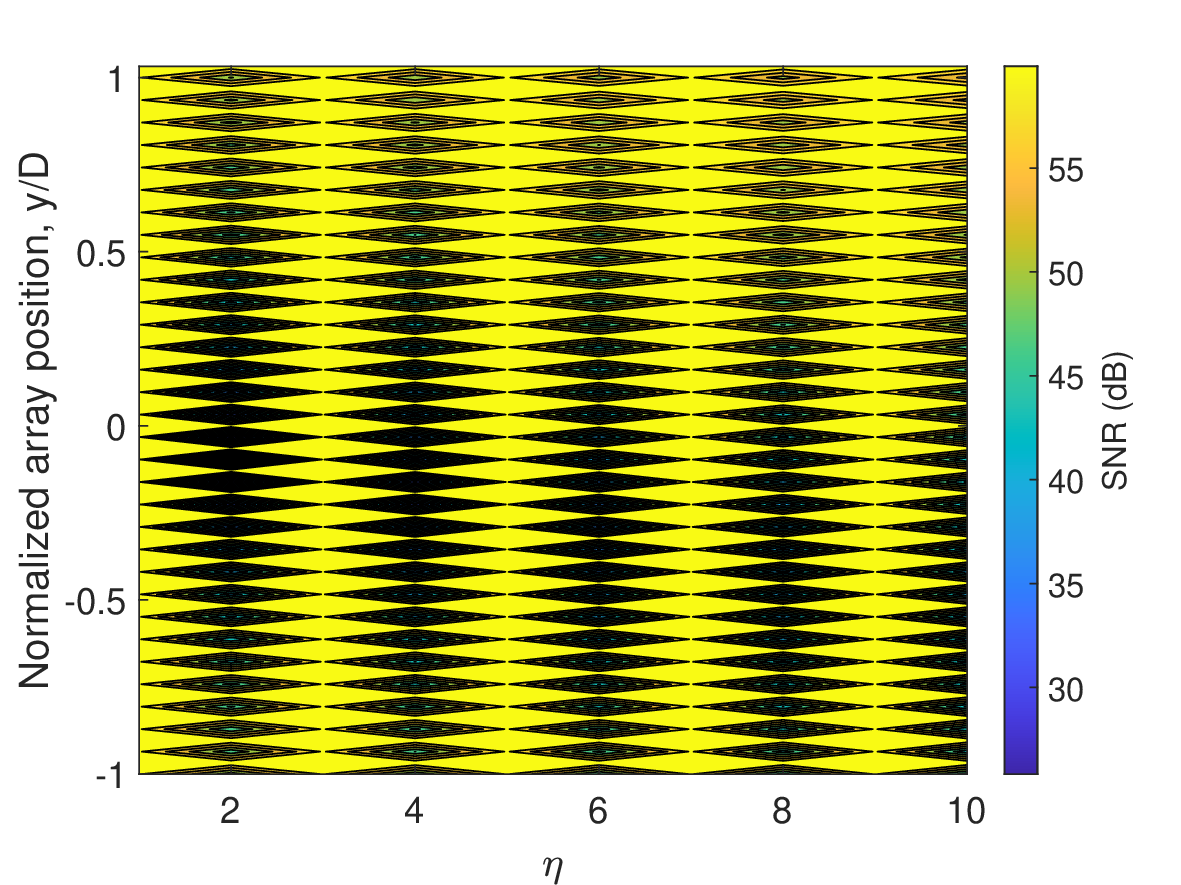}}
 \end{minipage}
 }
 \subfigure[Sum rate of the multi-user communication]{
 \begin{minipage}[t]{0.5\textwidth}
 \centering
 \centerline{\includegraphics[width=2.5in,height=1.875in]{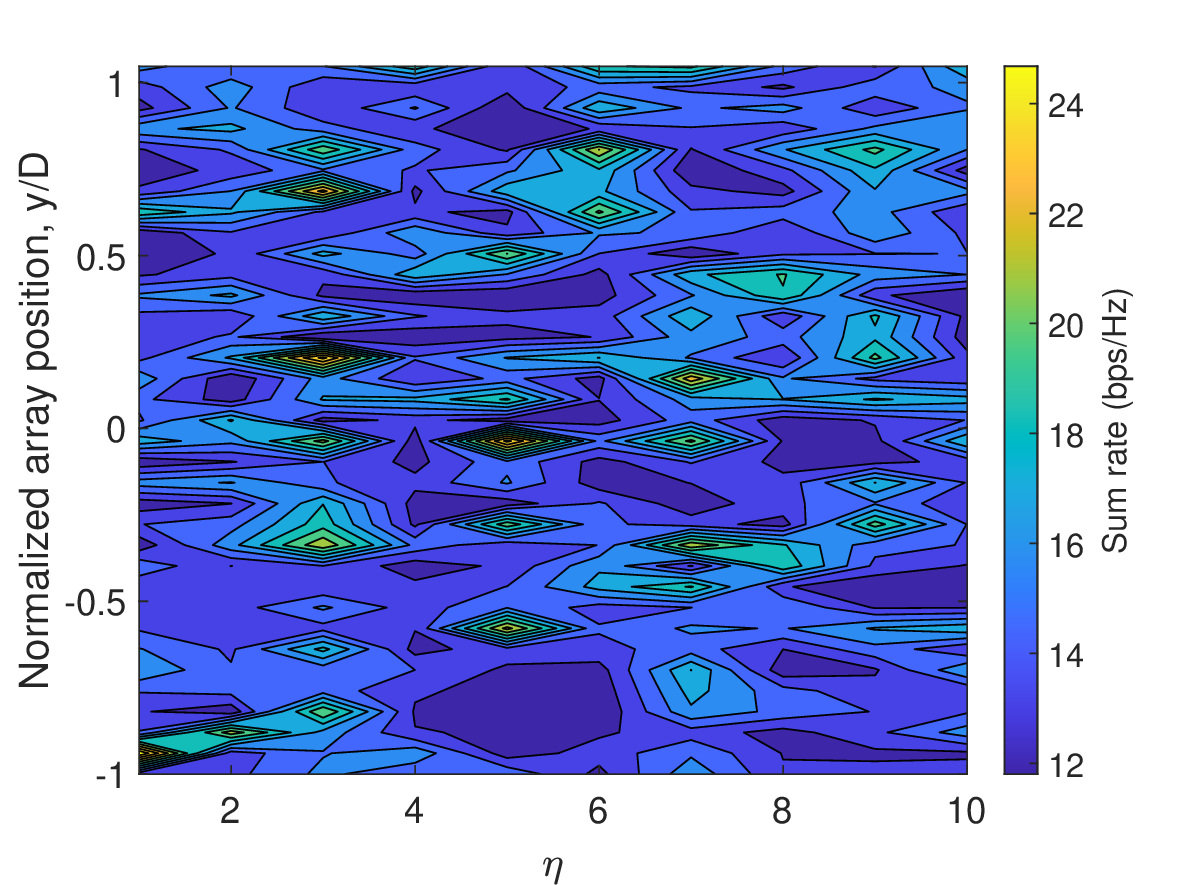}}
 \end{minipage}
 }
 \caption{Performance variations versus the normalized position and sparsity level of the GMA.}
 \label{fig:PerformanceVariationVersusNormalizedYAndEta}
 \vspace{-0.5cm}
 \end{figure}

 Fig.~\ref{fig:PerformanceVariationVersusNormalizedYAndEta} shows the performance variations versus the normalized position $y/D$ and sparsity level $\eta$ of the GMA for the single- and multi-user communications, respectively. For the single-user communication, two paths with equal amplitude is considered. It is firstly observed from Fig.~\ref{fig:PerformanceVariationVersusNormalizedYAndEta}(a) that the SNR of the single-user communication system exhibits significant variations as the array position and sparsity level change, and the performance gap between the maximum and minimum SNR exceeds 25 dB. Similarly, considerable variations of sum rate can be observed for the multi-user communication system in Fig.~\ref{fig:PerformanceVariationVersusNormalizedYAndEta}(b), and the performance gap between the maximum and minimum sum rates exceeds 10 bps/Hz in this example. The above results demonstrate the importance to optimize the position and sparsity level of GMA for single- and multi-user communications.

 \begin{figure}[!t]
 \setlength{\abovecaptionskip}{-0.15cm}
 \setlength{\belowcaptionskip}{-0.1cm}
 \centering
 \centerline{\includegraphics[width=2.5in,height=1.875in]{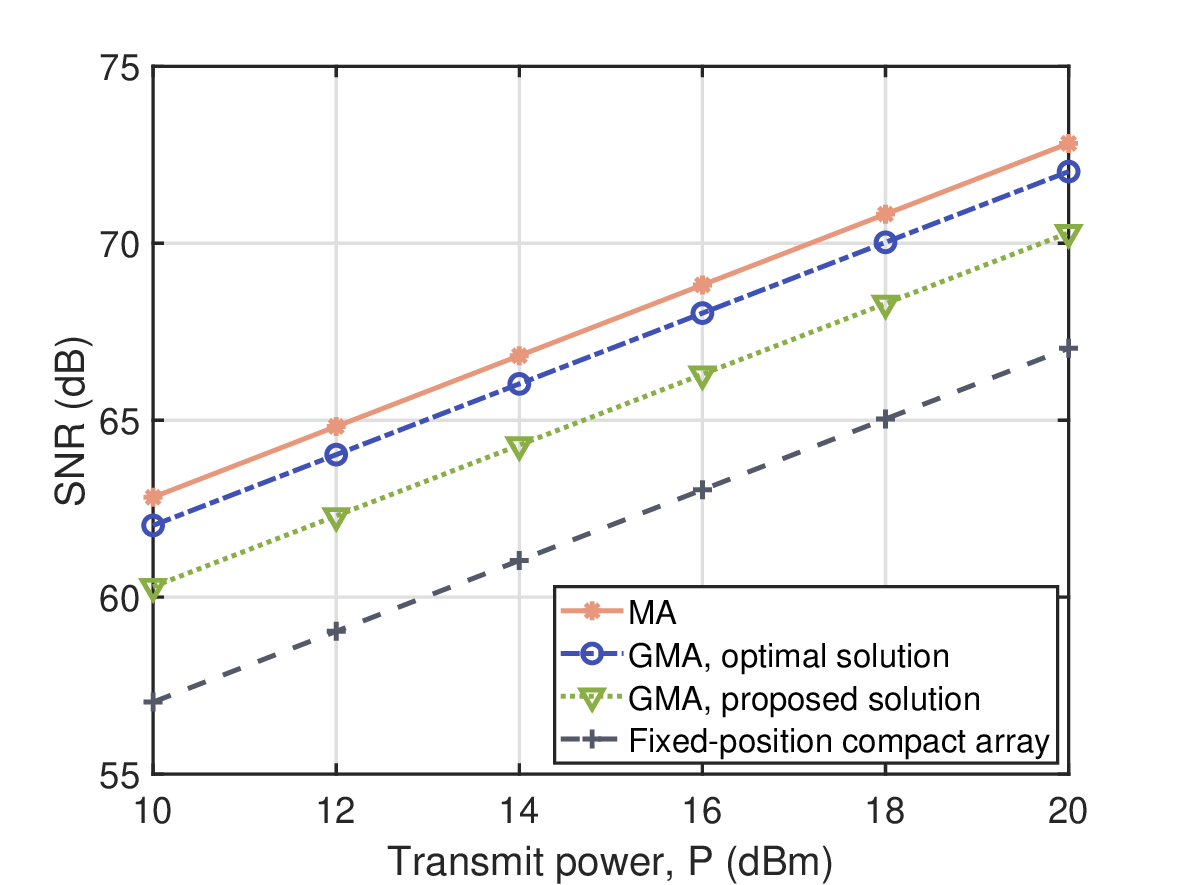}}
 \caption{SNR versus the transmit power.}
 \label{fig:SNRVersusTransmitPower}
 \vspace{-0.3cm}
 \end{figure}

 Fig.~\ref{fig:SNRVersusTransmitPower} shows the SNR versus the transmit power $P$ for the single-user communication system. The movable region size is set as $Y = 8D$, where $Y \triangleq  y_{\max} -y_{\min}$, and the number of physical array elements is $M= 128$. For comparison, two benchmarking schemes are considered: (1) MA: all the $N$ array elements can independently move subject to the minimum element spacing of $\lambda/2$; (2) Fixed-position compact array: the antenna array with $N$ elements separated by $\lambda/2$. Besides, the optimal solution to the GMA can be obtained via two-dimensional search. It is observed that though the GMA suffers from little performance loss as compared to MA, it significantly outperforms the conventional fixed-position compact array. This is expected since the GMA achieves a balance of mobility between MA and the fixed-position compact array. In particular, compared to MA, GMA not only avoids the prohibitive complexity for finding the optimal position of each antenna, but also eases the movement control.

 \begin{figure}[!t]
 \setlength{\abovecaptionskip}{-0.15cm}
 \setlength{\belowcaptionskip}{-0.15cm}
 \centering
 \centerline{\includegraphics[width=2.5in,height=1.875in]{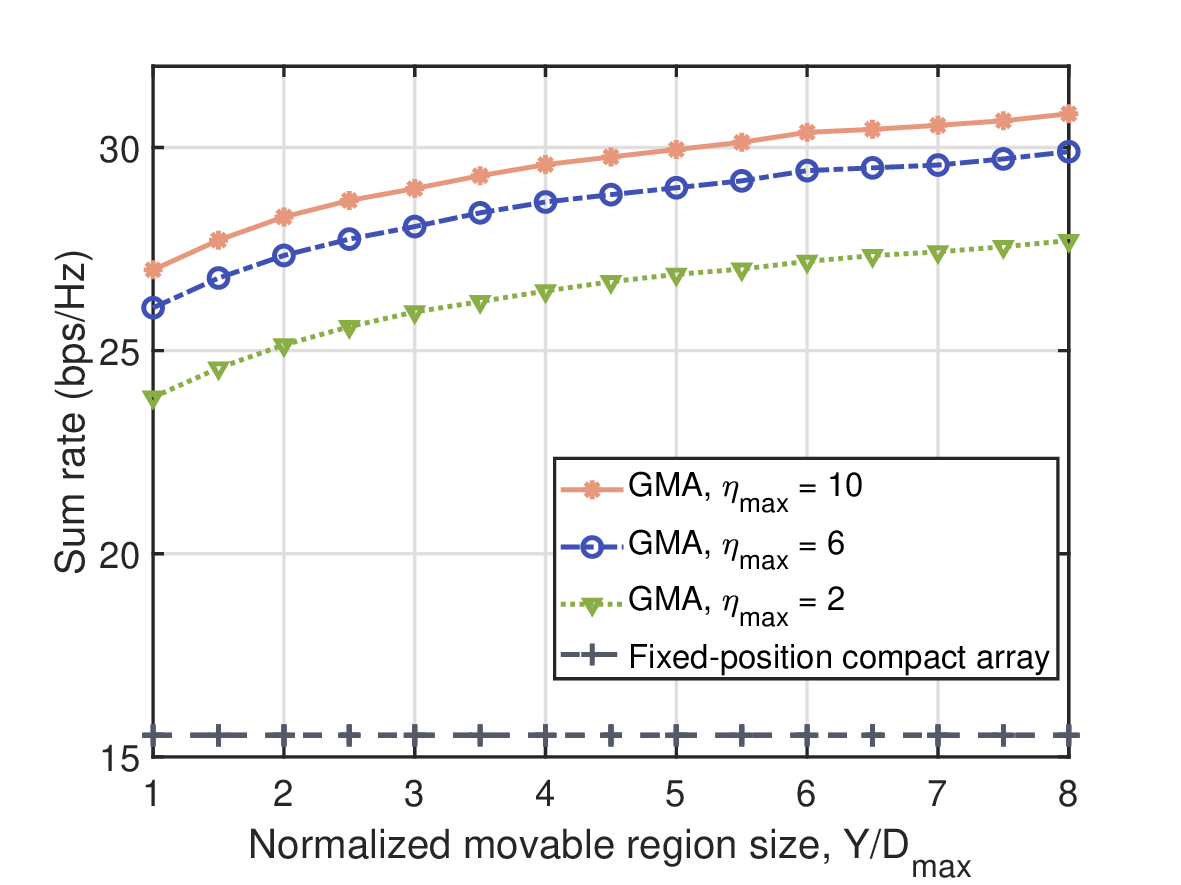}}
 \caption{Sum rate versus the normalized movable region size.}
 \label{fig:sumRateVersusNormalizedRegionSize}
 \vspace{-0.6cm}
 \end{figure}
 Fig.~\ref{fig:sumRateVersusNormalizedRegionSize} shows the sum rate versus the normalized movable region size $Y/D_{\max}$ for various maximum sparsity level $\eta_{\max}$, where $\eta_{\max} = \left\lfloor {(M - 1)/(N - 1)} \right\rfloor $ is varied by varying the number of physical array elements $M$, and $D_{\max} = 31d$, corresponding to the compact array dimension with $M =32$ elements. It is observed that the sum rate of the GMA significantly outperforms that of the fixed-position compact antenna, and the performance gain becomes more significant as the movable region size increases, which is expected since the GMA is likely to achieve higher interference mitigation gain with a larger movable region. Besides, the increase of $\eta_{\max}$ also contributes to the improvement of sum rate, since a larger sparsity can result in a larger array aperture, which also provides a more flexible array pattern to distinguish different paths of users in the spatial domain.

\vspace{-0.2cm}
\section{Conclusion}\label{sectionConclusion}
 This letter proposed a new architecture for MA termed GMA, where all the array elements move collectively and flexible array architecture is achieved by AS. By focusing on the uniform sparse array architecture with the antenna spacing determined by the sparsity level, we formulated the sum rate maximization problem by jointly optimizing the position and sparsity level of the GMA. An alternating optimization algorithm was then proposed to tackle the optimization problem. Numerical results verified the importance of proper movement and sparsity optimization for the GMA, and the significant performance gain achieved over the conventional FPA.

\bibliographystyle{IEEEtran}
\bibliography{refGroupMovableAntenna}

\end{document}